\documentclass[twoside,12pt]{article}
\usepackage{amssymb,amsmath,bm,euscript}
\usepackage{graphicx,color,caption}
\usepackage{ifpdf}

\usepackage[colorlinks=true]{hyperref}

\usepackage{braket}

\usepackage{algorithm, algorithmic}

\newtheorem{proposition}{Proposition}[section]
\newtheorem{theorem}[proposition]{Theorem}

\newcommand{\qed}{\hphantom{.}\hfill $\Box$\medbreak}

\def\A{{\mathcal{A}}}
\def\B{\mathcal{B}}

\def\x{{\bf x}}
\def\y{{\bf y}}

\def\0{{\bf 0}}

\title{\bf{Copositivity of Three-Dimensional
Symmetric Tensors}}
\author{ \hspace{1mm} Liqun Qi\thanks{
Department of Applied
    Mathematics, The Hong Kong Polytechnic University, Hung Hom,
    Kowloon, Hong Kong, China; ({\tt liqun.qi@polyu.edu.hk}).},
     \  \
   Yisheng Song\thanks{School of Mathematical Sciences, Chongqing University, Chongqing 401331 China; ({\tt yisheng.song@cqnu.edu.cn}). This author's work was supported by NSFC (Grant No.  11571095, 11601134). },
 \ and \
  Xinzhen Zhang\thanks{School of Mathematics, Tianjin University, Tianjin 300354 China; ({\tt xzzhang@tju.edu.cn}). This author's work was supported by NSFC (Grant No.  11871369). },
}

\begin{document}
\date{\today}
\maketitle

\begin{abstract}
In this paper, we seek analytically checkable necessary and sufficient condition for copositivity of a three-dimensional symmetric tensor.   We first show that for a general third order three-dimensional symmetric tensor, this means to solve a quartic equation and some quadratic equations.  All of them can be solved analytically.  Thus, we present an analytical way to check copositivity of a third order three dimensional symmetric tensor.   Then, we consider a model of vacuum stability for $\mathbb{Z}_3$ scalar dark matter.   This is a special fourth order three-dimensional symmetric tensor.  We show that an analytically expressed necessary and sufficient condition for this model bounded from below can be given, by using a result given by Ulrich and Watson in 1994.

\vskip 12pt \noindent {\bf Key words.} {copositive tensors, symmetric tensors, analytically checkable, vacuum stability.}

\vskip 12pt\noindent {\bf AMS subject classifications. }{15A69, 15A83}
\end{abstract}


\section{Introduction}

Checking that a scalar potential is bounded from below (BFB) is an ubiquitous and difficult task in particle physics.  For this task, copositivity of symmetric tensors plays an important role \cite{FI19, IKM18, Ka16, Ka18}.
Copositive tensors were introduced in 2013 \cite{Qi13}, and studied in \cite{SQ15}.   Motivated by the BFB study in physics  \cite{FI19, IKM18, Ka16, Ka18}, and the tensor complementarity problem study in optimization \cite{QCC18, QL17}, various testing methods for detecting if a symmetric tensor is copositive or not appeared \cite{CHQ17, CHQ18, CW18, LZHQ19, NYZ18}.   Recently, some analytical expressable sufficient conditions for copositivity of third order and fourth order three-dimensional symmetric tensors
also appeared \cite{LS19, SQ20}.    However, it is still very difficult to find analytically expressable
necessary and sufficient conditions for copositivity of third order and fourth order three-dimensional symmetric tensors, while such conditions are very useful in particle physics \cite{FI19, IKM18, Ka16, Ka18}.

In this paper, we seek analytically checkable necessary and sufficient condition for copositivity of an three-dimensional symmetric tensor for two problems.   The first problem is to check copositivity
for a general third order three-dimensional symmetric tensor.   The second problem is  a model of vacuum stability for $\mathbb{Z}_3$ scalar dark matter, studied in \cite{Ka16}.  This is to check copositivity of a special fourth order three-dimensional symmetric tensor.  A theorem of Ulrich and Watson \cite{UW94} is used for solve the second problem.


In the next section, preliminary knowledge on copositive tensors is presented.  A theorem of Ulrich and Watson \cite{UW94} in 1994 is also stated there.    In Section 3, we present a necessary and sufficient condition for copositivity of an $m$th order three-dimensional symmetric tensor.   To check this condition, a one variable polynomial equation of degree $(m-1)^2$, and some one variable polynomial equations of degree $m-1$, need to be solved.
In Section 4, we present a necessary and sufficient condition for copositivity of a third order three-dimensional symmetric tensor.    By using a theorem given in \cite{LS19}, this result is simpler than the general case.    The works need to be done is to solve a quartic equation and at most four quadratic equations.   We may solve them analytically.
We then present an analytically expressed necessary and sufficient condition for the vacuum stability model for $\mathbb{Z}_3$ scalar dark matter in Section 5.

\section{Preliminaries}

We denote the set of all $m$th order $n$-dimensional real symmetric tensors by $S_{m, n}$, where $m$ and $n$ are positive integers, $m, n \ge 2$.      For $\A = (a_{i_1\cdots i_m}) \in S_{m, n}$, we have
$i_1, \cdots, i_m  = 1, \cdots, n$ and $a_{i_1\cdots i_m}$ is invariant under any index permutation.
For $\x = (x_1, \cdots, x_n)^\top \in \Re^n$,
$$\A \x^m : = \sum_{i_1, \cdots, i_m = 1}^n a_{i_1\cdots i_m}x_{i_1}\cdots x_{i_m}.$$
We say that $\A$ is copositive if for any $\x \in \Re_{+}^n$, we have
$$\A \x^m \ge 0.$$
We say that $\A$ is  strictly copositive if for any $\x \in \Re_{+}^n$, $\x \not = \0$, we have
$$\A \x^m > 0.$$

We have
$${\partial \over \partial x_i}\left({1 \over m}\A \x^m\right) = \left(\A \x^{m-1}\right)_i =
\sum_{i_2, \cdots, i_m = 1}^n a_{ii_2\cdots i_m}x_{i_2}\cdots x_{i_m}.$$

Liu and Song \cite{LS19} proved the following theorem.   We will use this theorem in Section 4. 

\begin{theorem} (\cite[Theorem 3.1]{LS19}) \label{t2.1}
Suppose that $\B = (b_{ijk}) \in S_{3, 2}$.   Then $\B$ is copositive if and only if $b_{111} \ge 0$, $b_{222} \ge 0$, and either 

(a) $b_{112} \ge 0$, $b_{122} \ge 0$; or

(b) $\max \{ b_{111}, b_{222} \} > 0$, 
$$b_{111}b_{122}^3 + 4 b_{112}^3b_{222} + b_{111}^2b_{222}^2 - 6b_{111}b_{112}b_{122}b_{222} - 3b_{112}^2
b_{122}^2 \ge 0.$$
\end{theorem}

For a quartic polynomial $g(t)$ with real coefficients,
\begin{equation}\label{eq:26}g(t)=at^4+bt^3+ct^2+dt+e,\end{equation}
Ulrich and Watson \cite{UW94} proved the following theorem.   We will use this theorem in Section 5.

\begin{theorem} (\cite[Theorem 2]{UW94}) \label{le:24} Let $g(t)$ be a quartic and univariate polynomial defined by \eqref{eq:26} with $a>0$ and $e>0$. Define
$$\begin{aligned}
\alpha &=ba^{-\frac34}e^{-\frac14},\ \beta=ca^{-\frac12}e^{-\frac12},\ \gamma=da^{-\frac14}e^{-\frac34},\\
\Delta&=4(\beta^2-3\alpha\gamma+12)^3-(72\beta+9\alpha\beta\gamma-2\beta^3-27\alpha^2-27\gamma^2)^2,\\
\mu &=(\alpha-\gamma)^2-16(\alpha+\beta+\gamma+2),\\
\eta&=(\alpha-\gamma)^2-\frac{4(\beta+2)}{\sqrt{\beta-2}}(\alpha+\gamma+4\sqrt{\beta-2}).
\end{aligned}$$
Then (i) $g(t)\geq0$ for all $t>0$ if and only if
\begin{itemize}
	\item[(1)]\ \ \ $\beta<-2\ \mbox{ and }\ \Delta \leq0\ \mbox{ and }\ \alpha+\gamma>0;$
	\item[(2)]\ \ \ $-2\leq \beta\leq 6\ \mbox{ and } \begin{cases} \Delta\leq 0\ &\mbox{ and }\ \alpha+\gamma>0\\ & or \\ \Delta\geq 0\ &\mbox{ and }\ \mu\leq 0\end{cases}$\\
	\item[(3)]\ \ \ $\beta>6\ \mbox{ and } \begin{cases} \Delta\leq 0\ &\mbox{ and }\ \alpha+\gamma>0\\ & or \\ \alpha>0\ &\mbox{ and }\ \gamma>0\\ & or  \\ \Delta\geq 0\ &\mbox{ and }\eta\leq 0.\end{cases}$
\end{itemize}
(ii) $g(t)\geq0$ for all $t>0$ if
\begin{itemize}
	\item[(1)]\ \ \ \ $\alpha>-\frac{\beta+2}{2} \mbox{ and }\gamma>-\frac{\beta+2}{2} \mbox{ for }\beta\leq6;$
\item[(2)]\ \ \ \ $\alpha>-2\sqrt{\beta-2} \mbox{ and }\gamma>-2\sqrt{\beta-2} \mbox{ for }\beta>6.$
\end{itemize}
\end{theorem}

\section{A Necessary and Sufficient Condition}

Let $A = (1, 0, 0)$, $B = (0, 1, 0)$ and $C = (0, 0, 1)$.   Denote the triangle $\Delta ABC$ by
$S \equiv \Delta ABC = \{ \x \in \Re^3 : x_1+x_2+x_3 = 1, x_1 \ge 0, x_2 \ge 0, x_3 \ge 0 \}$.  The three edges of $S$ are $AB, BC$ and $CA$.

For a three-dimensional symmetric tensor $\A$, we have the following necessary and sufficient condition for its copositivity.

\begin{theorem} \label{t3.1}
Suppose that $\A = (a_{i_1\cdots i_m}) \in S_{m, 3}$, where the integer $m \ge 2$.  
Denote 
$$\psi_i(x_1, x_2, x_3) = \sum_{i_2, \cdots, i_m = 1}^3 a_{ii_2\cdots i_m}x_{i_2}\cdots x_{i_m},$$
for $i = 1, 2, 3$.
Then $\A$ is copositive if and only if the following five conditions are satisfied:

(1) $a_{1\cdots 1} \ge 0$, $a_{2\cdots 2} \ge 0$, and $a_{3\cdots 3} \ge 0$;

(2) There are no $x_1 > 0$ and $x_2 > 0$ such that
\begin{equation} \label{e3.1}
\sum_{i_2, \cdots, i_m = 1, 2} a_{1i_2\cdots i_m}x_{i_2}\cdots x_{i_m} = \sum_{i_2, \cdots, i_m = 1, 2} a_{2i_2\cdots i_m}x_{i_2}\cdots x_{i_m} < 0,\ x_1+x_2 = 1;
\end{equation}

(3) There are no $x_1 > 0$ and $x_3 > 0$ such that
\begin{equation} \label{e3.2}
\sum_{i_2, \cdots, i_m = 1, 3} a_{1i_2\cdots i_m}x_{i_2}\cdots x_{i_m} = \sum_{i_2, \cdots, i_m = 1, 3} a_{3i_2\cdots i_m}x_{i_2}\cdots x_{i_m} < 0,\ x_1+x_3 = 1;
\end{equation}

(4) There are no $x_2 > 0$ and $x_3 > 0$ such that
\begin{equation} \label{e3.3}
\sum_{i_2, \cdots, i_m = 2, 3} a_{2i_2\cdots i_m}x_{i_2}\cdots x_{i_m} = \sum_{i_2, \cdots, i_m = 2, 3} a_{3i_2\cdots i_m}x_{i_2}\cdots x_{i_m} < 0,\ x_2+x_3 = 1;
\end{equation}

(5) There are no $y_1 > 0$ and $y_2 > 0$ such that
\begin{equation} \label{e3.4}
\psi_1(y_1, y_2, 1) = \psi_2(y_1, y_2, 1) = \psi_3(y_1, y_2, 1) < 0.
\end{equation}

If the three ``$\ge$'' inequalities in (1) are changed to the ``$>$'' inequalities, and the four ``$<$'' inequalities in (\ref{e3.1}-\ref{e3.4}) are changed to the ``$\le$'' inequalities, then we have a necessary and sufficient condition for strict copositivity.
\end{theorem}
{\bf Proof} Clearly, $\A$ is copositive if and only if for all $\y \in S$, $\A \y^m \ge 0$, i.e.,

(a) $\A \y^m \ge 0$ if $\y$ is one of the vertices $A$, $B$ and $C$;

(b) $\A \y^m \ge 0$ if $\y$ is in the relative interior of the edge $AB$;

(c) $\A \y^m \ge 0$ if $\y$ is in the relative interior of the edge $CA$;

(d) $\A \y^m \ge 0$ if $\y$ is in the relative interior of the edge $BC$;

(e)  $\A \y^m \ge 0$ if $\y$ is in the relative interior of $S$.

Clearly, condition (a) is equivalent to condition (1).

Condition (b) does not hold if and only if there is a global minimizer $(x_1, x_2)$ of the following minimization problem
\begin{equation} \label{e3.6}
\min \left\{ {1 \over m}\sum_{i_1, \cdots, i_m =1}^2 a_{i_1\cdots i_m}y_{i_1}\cdots y_{i_m} : y_1 + y_2 = 1, y_1 \ge 0, y_2 \ge 0 \right\},
\end{equation}
such that $x_1 > 0$, $x_2 > 0$ and the global minimum of (\ref{e3.6}) at $(x_1, x_2)$ is negative.
By the optimality conditions of (\ref{e3.6}), we have
\begin{eqnarray*}
\sum_{i_2, \cdots, i_m = 1, 2} a_{ii_2\cdots i_m}x_{i_2}\cdots x_{i_m} - \lambda & = & \mu_i, \ {\rm for} \ i = 1, 2, \\
x_1 + x_2 & = & 1, \\
x_1 \ge 0, & & x_2 \ge 0, \\
\mu_1 \ge 0, & & \mu_2 \ge 0, \\
x_i\mu_i & = & 0, \ {\rm for} \ i = 1, 2,
\end{eqnarray*}
where $\lambda$, $\mu_1$ and $\mu_2$ are Langrangian multipliers.   Since $x_1 > 0$ and $x_2 > 0$,  we have $\mu_1 = \mu_2 = 0$.  Thus,
\begin{eqnarray*}
\sum_{i_2, \cdots, i_m = 1, 2} a_{ii_2\cdots i_m}x_{i_2}\cdots x_{i_m} & = & \lambda, \ {\rm for} \ i = 1, 2, \\
x_1 + x_2 & = & 1, \\
x_1 \ge 0, x_2 \ge 0, & & x_3 = 0.
\end{eqnarray*}
Then
$$\lambda = \sum_{i=1}^3 x_i \sum_{i_2, \cdots, i_m = 1, 2} a_{ii_2\cdots i_m}x_{i_2}\cdots x_{i_m} < 0.$$
This shows that conditions (b) and (2) are equivalent.   Similarly, conditions (c) and (3) are equivalent;
conditions (d) and (4) are equivalent.

Condition (e) does not hold if and only if there is a minimizer $(x_1, x_2, x_3)$ of the following minimization problem
\begin{equation} \label{e3.7}
\min \left\{ {1 \over m}\A \y^m : y_1 + y_2 + y_3 = 1, y_1 \ge 0, y_2 \ge 0, y_3 \ge 0 \right\},
\end{equation}
$x_1 > 0$, $x_2 > 0$ and $x_3 > 0$, such that the minimum value is negative.
By the optimality conditions of (\ref{e3.7}), we have
\begin{eqnarray*}
\psi_i(x_1, x_2, x_3) - \lambda & = & \mu_i, \ {\rm for} \ i = 1, 2, 3, \\
x_1 + x_2 + x_3 & = & 1, \\
x_i & \ge & 0, \ {\rm for} \ i = 1, 2, 3, \\
\mu_i & \ge & 0, \ {\rm for} \ i = 1, 2, 3, \\
x_i\mu_i & = & 0, \ {\rm for} \ i = 1, 2, 3,
\end{eqnarray*}
where $\lambda$, $\mu_1, \mu_2$ and $\mu_3$ are Langrangian multipliers.   Since $x_1 > 0$, $x_2 > 0$ and $x_3 > 0$, we have $\mu_1 = \mu_2 = \mu_3 = 0$.  Hence,
\begin{eqnarray*}
\psi_i(x_1, x_2, x_3) & = & \lambda, \ {\rm for} \ i = 1, 2, 3, \\
x_1 + x_2 + x_3 & = & 1, \\
x_i & > & 0,  \ {\rm for} \ i = 1, 2, 3.
\end{eqnarray*}
Then
$$\lambda = \sum_{i=1}^3 x_i \psi_i(x_1, x_2, x_3) = \A \x^m < 0.$$
Let $x_1 = y_1x_3$ and $x_2 = y_2x_3$.
We see that conditions (e) and (5) are equivalent.

The extension to strict copositivity is clear.
\qed

Condition (1) is very easy to check.

Consider condition (2).  Substitute $x_2 = 1 - x_1$ to (\ref{e3.1}).  Let
$$\phi_1(x_1) = \sum_{i_2, \cdots, i_m = 1, 2} a_{1i_2\cdots i_m}x_{i_2}\cdots x_{i_m},$$
$$\phi_2(x_1) = \sum_{i_2, \cdots, i_m = 1, 2} a_{2i_2\cdots i_m}x_{i_2}\cdots x_{i_m}$$
and
$$\phi(x_1) = \phi_1(x_1) - \phi_2(x_1).$$
Then, checking if condition (2) holds is equivalent to solve the one-dimensional polynomial equation
$$\phi(x_1) = 0,$$
where $\phi$ is a polynomial of $x_1$, with degree $m-1$, to confirm that $\phi$ has no root $x_1$ such that
$0 < x_1 < 1$ and $\phi_1(x_1) < 0$.

Conditions (3) and (4) can be checked similarly.


We now study the procedure to check condition (5) of Theorem \ref{t3.1}.

Let
$$\psi_4(y_1, y_2) = \psi_1(y_1, y_2, 1) - \psi_2(y_1, y_2, 1),\ \psi_5(y_1, y_2) = \psi_1(y_1, y_2, 1) - \psi_3(y_1, y_2, 1).$$
Then, checking if condition (5) holds is equivalent to solve the system of polynomial equations
\begin{equation} \label{e4.8}
\psi_4(y_1, y_2) = 0,\ \psi_5(y_1, y_2) = 0,
\end{equation}
where $\psi_4$ and $\psi_5$ are polynomials of $y_1$ and $y_2$, with degree $m-1$, to confirm that (\ref{e4.8}) has no solution $(y_1, y_2)$ such that
$y_1 > 0$, $y_2 > 0$ and $\psi_1(y_1, y_2, 1) < 0$.

To solve the system (\ref{e4.8}), we may first regard it as a system of polynomial equations of $y_2$
\begin{equation} \label{e4.9}
\sum_{i=0}^{m-1} \eta_iy_2^{m-i-1} = 0,  \ \sum_{i=0}^{m-1} \tau_iy_2^{m-i-1} = 0,
\end{equation}
where $\eta_i$ and $\tau_i$ are polynomials of $y_1$ with degree $i$, and can be calculated by (\ref{e4.8}), for $i = 0, \cdots m-1$.   By the Sylvester theorem, system (\ref{e4.9}) has a solution if and only if its resultant vanishes \cite{GKZ94}. The resultant of (\ref{e4.9}) is a $2(m-1) \times 2(m-1)$ determinant
$$G(y_1) = \left|\begin{array}{ccccccccc}
\eta_0 & \eta_1 & \cdots & \eta_{m-2} & \eta_{m-1} & 0 & \cdots & 0 & 0 \\
0 & \eta_0 & \cdots & \eta_{m-3} & \eta_{m-2} & \eta_{m-1} & \cdots & 0 & 0 \\
\cdot & \cdot & \cdots & \cdot & \cdot & \cdot & \cdots & \cdot & \cdot \\
0 & 0 & \cdots & \eta_0 & \eta_1 & \eta_2 & \cdots & \eta_{m-1} & 0 \\
0 & 0 & \cdots & 0 & \eta_0 & \eta_1 & \cdots & \eta_{m-2} & \eta_{m-1} \\
\tau_0 & \tau_1 & \cdots & \tau_{m-2} & \tau_{m-1} & 0 & \cdots & 0 & 0 \\
0 & \tau_0 & \cdots & \tau_{m-3} & \tau_{m-2} & \tau_{m-1} & \cdots & 0 & 0 \\
\cdot & \cdot & \cdots & \cdot & \cdot & \cdot & \cdots & \cdot & \cdot \\
0 & 0 & \cdots & \tau_0 & \tau_1 & \tau_2 & \cdots & \tau_{m-1} & 0 \\
0 & 0 & \cdots & 0 & \tau_0 & \tau_1 & \cdots & \tau_{m-2} & \tau_{m-1}
\end{array}\right|,$$
which is a polynomial of $y_1$ with degree $(m-1)^2$.   Find all of its roots satisfying
$y_1 > 0$.   Substitute such roots to
\begin{equation} \label{e4.10}
\sum_{i=0}^{m-1} \eta_iy_2^{m-i} = 0.
\end{equation}
For each root $\alpha$, we have a polynomial equation of $y_2$ with degree $m-1$.   Find all of its positive solutions $\beta >0$, where $\alpha$ is the corresponding root of $G$.  For all such solution pairs $(\alpha, \beta)$,
check if $\psi_1(\alpha, \beta, 1) = \psi_2(y_1, y_2, 1) = \psi_3(y_1, y_2, 1) < 0$ or not.  If there is such a solution, then condition (5) of Theorem \ref{t3.1} is violated.  Otherwise,  condition (5) of Theorem \ref{t3.1} is satisfied.

Hence, for checking condition (5) of Theorem \ref{t3.1}, we need to solve a polynomial equation of degree $(m-1)^2$ and at most $(m-1)^2$ polynomial equation of degree $m-1$.   Totally, for checking conditions of Theorem \ref{t3.1}, we need to solve a polynomial equation of degree $(m-1)^2$, and at most $(m-1)^2 +3$ polynomial equations of degree $m-1$.

In particular, for checking copositivity of a third order three-dimensional symmetric tensor, we only need to solve a quartic equation and at most seven quadratic equations.   These can be done analytically.  We will study this in the next section.




\section{Third Order Three-Dimensional Symmetric Tensors}

Suppose that $\A = (a_{ijk}) \in S_{3, 3}$.   Then $\A$ has ten independent entries $a_{111}$, $a_{222}$, $a_{333}$, $a_{112} = a_{121} = a_{211}$, $a_{122} = a_{212} = a_{221}$, $a_{113} = a_{131} = a_{311}$,
$a_{133} = a_{313} = a_{331}$, $a_{223} = a_{232} = a_{322}$, $a_{233} = a_{323} = a_{332}$, and $a_{123}
= a_{231} = a_{312} = a_{213} = a_{132} = a_{321}$.

By using Theorem \ref{t2.1}, the following theorem is simpler than Theorem \ref{t3.1} with $m = 3$.  

\begin{theorem} \label{t4.1}
Suppose that $\A = (a_{ijk}) \in S_{3, 3}$.
Denote
$$\psi_i(x_1, x_2, x_3) = \sum_{j, k = 1}^3 a_{ijk}x_jx_k,$$
for $i = 1, 2, 3$.
Then $\A$ is copositive if and only if the following five conditions are satisfied:

(1) $a_{111} \ge 0$, $a_{222} \ge 0$, $a_{333} \ge 0$; 

(2) either $a_{112} \ge 0$ and $a_{122} \ge 0$, or $\max \{ a_{111}, a_{222} \} > 0$ and
$$a_{111}a_{122}^3 + 4 a_{112}^3a_{222} + a_{111}^2a_{222}^2 - 6a_{111}a_{112}a_{122}a_{222} - 3a_{112}^2
a_{122}^2 \ge 0;$$

(3) either $a_{113} \ge 0$ and $a_{133} \ge 0$, or $\max \{ a_{111}, a_{333} \} > 0$ and
$$a_{111}a_{133}^3 + 4 a_{113}^3a_{333} + a_{111}^2a_{333}^2 - 6a_{111}a_{113}a_{133}a_{333} - 3a_{113}^2
a_{133}^2 \ge 0;$$

(4) either $a_{223} \ge 0$ and $a_{233} \ge 0$, or $\max \{ a_{222}, a_{333} \} > 0$ and
$$a_{222}a_{233}^3 + 4 a_{223}^3a_{333} + a_{222}^2a_{333}^2 - 6a_{222}a_{223}a_{233}a_{333} - 3a_{223}^2
a_{233}^2 \ge 0;$$

(5) There are no $y_1 > 0$ and $y_2 > 0$ such that
\begin{equation} \label{e4.11}
\psi_1(y_1, y_2, 1) = \psi_2(y_1, y_2, 1) = \psi_3(y_1, y_2, 1) < 0.
\end{equation}
\end{theorem} 
{\bf Proof} Conditions (1) and (5) are from Theorem \ref{t3.1}.   By applying Theorem \ref{t2.1} to the three edges of $\Delta ABC$, we have conditions (2), (3) and (4).
\qed

Not only condition (1), but also conditions (2), (3) and (4) are explicitly given.   Thus, Theorem \ref{t4.1} is simpler than Theorem \ref{t3.1} with $m = 3$. 

As to condition (5) of Theorem \ref{t4.1}, for $i = 1, 2, 3$, we have
$$\psi_i(y_1, y_2, 1) = a_{i11}y_1^2 + a_{i22}y_2^2 + 2a_{i12}y_1y_2 + 2a_{i13}y_1 + 2a_{i23}y_2 + a_{i33}.$$
Then,
$$\psi_4(y_1, y_2) = \eta_0y_2^2 +\eta_1y_2 + \eta_2,\
\psi_5(y_1, y_2) = \tau_0y_2^2 + \tau_1y_2 + \tau_2.$$
We have
$$\eta_0 = a_{122} - a_{222},$$
$$\eta_1 = 2\left[(a_{112}-a_{122})y_1 +a_{123} - a_{223}\right],$$
$$\eta_2 =  (a_{111}-a_{112})y_1^2 + 2(a_{113}-a_{123})y_1+a_{133}-a_{233},$$
$$\tau_0 = a_{122}- a_{223},$$
$$\tau_1 = 2\left[(a_{112}-a_{123})y_1 +a_{123} - a_{233}\right],$$
$$\tau_2 = (a_{111}-a_{113})y_1^2+2(a_{112}-a_{123})y_1 +a_{133}- a_{333}.$$
Then
$$G(y_1) = \left|\begin{array} {cccc} \eta_0 & \eta_1 & \eta_2 & 0 \\
0 & \eta_0 & \eta_1 & \eta_2 \\
\tau_0 & \tau_1 & \tau_2 & 0 \\
0 & \tau_0 & \tau_1 & \tau_2
\end{array}\right| = \eta_0 \left|\begin{array} {ccc} \eta_0 & \eta_1 & \eta_2 \\
\tau_1 & \tau_2 & 0 \\ \tau_0 & \tau_1 & \tau_2
\end{array}\right| + \tau_0\left|\begin{array} {ccc} \eta_1 & \eta_2 & 0 \\ \eta_0 & \eta_1 & \eta_2 \\
\tau_0 & \tau_1 & \tau_2
\end{array}\right|
$$
$$= \eta_0^2\tau_2^2 + \eta_0\eta_2\tau_1^2 - 2\eta_0\eta_2\tau_0\tau_2 - \eta_0\eta_1\tau_1\tau_2 + \eta_1^2\tau_0\tau_2 + \eta_2^2\tau_0^2 - \eta_1\eta_2\tau_0\tau_1.$$
Thus, $G(y_1)=0$ is a quartic equation of $y_1$.   We may write
$$G(y_1) = ay_1^4 + by_1^3 + cy_1^2 + dy_1 + e.$$

If $a=b=0$, then $G(y_1) = 0$ is a quadratic equation, or a linear equation, or a constant equation.  It is easy to find its real roots.

If $a=0$ but $b \not = 0$, then $G(y_1) = 0$ is a cubic equation. Let $y_1 = z - {c \over 3b}$.  Then we may convert $G(y_1) = 0$ to its depressed form
\begin{equation} \label{e5.11}
z^3 + pz + q = 0.
\end{equation}
The discriminant of (\ref{e5.11}) is
$$\Delta = -4p^3 -27q^2.$$
By Cardano's formula, (\ref{e5.11}) always has one real root:
$$\sqrt[3]{-{q \over 2} + \sqrt{{q^2 \over 4} + {p^3 \over 27}}} + \sqrt[3]{-{q \over 2} - \sqrt{{q^2 \over 4} + {p^3 \over 27}}}.$$

If $\Delta \ge 0$, then (\ref{e5.11}) has two more real roots (maybe multiple):
$$\xi\sqrt[3]{-{q \over 2} + \sqrt{{q^2 \over 4} + {p^3 \over 27}}} + \xi^2\sqrt[3]{-{q \over 2} - \sqrt{{q^2 \over 4} + {p^3 \over 27}}}$$
and
$$\xi^2\sqrt[3]{-{q \over 2} + \sqrt{{q^2 \over 4} + {p^3 \over 27}}} + \xi\sqrt[3]{-{q \over 2} - \sqrt{{q^2 \over 4} + {p^3 \over 27}}},$$
where
$$\xi= {-1+\sqrt{-3} \over 2}, \ \xi^2 = {-1-\sqrt{-3} \over 2}.$$
From these, we have the roots of $G(y_1) = 0$ by $y_1 = z - {c \over 3b}$.

If $a \not = 0$, then by letting $y_1 = \sqrt[4]{a}\left(z - {b \over 4a}\right)$, we convert $G(y_1)=0$ to its depressed form
\begin{equation} \label{e5.12}
z^4 + pz^2 + qy + r = 0.
\end{equation}
Here, $p$ and $q$ are different from $p$ and $q$ used before.  If $p=q=r=0$, (\ref{e5.12}) is trivial to solve.   Assume that it is not in this case.   Then we may use Descartes' method in 1637 to factor $z^4 + pz^2+qz+r$ \cite{De54}.   Suppose that
 \begin{equation} \label{e5.13}
z^4 + pz^2 + qz + r = (z^2 -uz+t)(z^2+uz+v)
\end{equation}
and $U = u^2$.  Then we have the resolvent cubic of (\ref{e5.12}):
 \begin{equation} \label{e5.14}
U^3 + 2pU^2 + (p^2-4r)U - q^2.
\end{equation}
Using the Cardano's formula described above, we may find the roots of the resolvent cubic.  Because of our assumption, at least one root is nonzero.  Taking square root of such a nonzero root of the resolvent cubic, we find the solution $u$ in (\ref{e5.13}).   Then we have
$$t = {1 \over 2}\left(p+u^2+{q \over u}\right),\ v = {1 \over 2}\left(p+u^2-{q \over u}\right).$$
With such a factorization (\ref{e5.13}) and $y_1 = \sqrt[4]{a}\left(z - {b \over 4a}\right)$, we find four roots of $G(y_1) = 0$.

For any real positive root $y_1 = \alpha$ of $G(y_1) = 0$, substitute it to $\eta_1$ and $\eta_2$.   Then solve $\psi_4(\alpha, y_2) = 0$ to find its root.  If $\psi_4(\alpha, y_2) = 0$ has a real positive solution $\beta$, then check if $\psi_1(\alpha, \beta, 1) = \psi_2(\alpha, \beta, 1)= \psi(\alpha, \beta, 1) < 0$ or not.  If so, then condition (5) of Theorem \ref{t4.1} is violated.   If no such pair $(\alpha, \beta)$ exists, then condition (5) of Theorem \ref{t4.1} is satisfied.

In this way, we have an analytical way to check if $\A$ is copositive or not.

\section{Vacuum Stability for $\mathbb{Z}_3$ Scalar Dark Matter}

Let $\B = (b_{ijkl}) \in S_{4, 3}$ be a general fourth order three-dimensional symmetric tensor.   Then $\B$ has fifteen independent entries $b_{1111}$, $b_{2222}$, $b_{3333}$, $b_{1112}$,  $b_{1113}$, $b_{1222}$, $b_{2223}$,  $b_{1333}$, $b_{2333}$, $b_{1122}$,  $b_{1133}$, $b_{2233}$, $b_{1123}$,
$b_{1223}$  and $b_{1233}$.

According to \cite{Ka16}, the most general scalar potential of the Standard Model Higgs $H_1$, an inert doublet $H_2$ and a complex $S$ which is symmetric under a $\mathbb{Z}_3$ group can be expressed as a quartic form
\begin{equation} \label{e6.15}
f(h_1, h_2, s) = \lambda_1h_1^4 + \lambda_2h_2^4 + \lambda_3h_1^2h_2^2 + \lambda_4\rho^2h_1^2h_2^2 +\lambda_Ss^4 + \lambda_{S1}s^2h_1^2 + \lambda_{S2}s^2h_2^2 - |\lambda_{S12}|\rho s^2h_1h_2,
\end{equation}
where $h_1, h_2$ and $s$ are physical quantities related with $H_1$, $H_2$ and $S$, $\lambda_1, \lambda_2, \lambda_3, \lambda_4, \lambda_S, \lambda_{S1}, \lambda_{S2}, \lambda_{S12}$ and $\rho$ are physical parameters, $0 \le \rho \le 1$.   See (89) of \cite{Ka16} for their meanings.

Let $x_1 = h_1$, $x_2 = h_2$, $x_3 = s$, $b_{1111} = \lambda_1$, $b_{2222} = \lambda_2$, $b_{3333} = \lambda_S$, $b_{1112} = b_{1113} = b_{1222} = b_{2223} = b_{1333} = b_{2333} = 0$, $b_{1122} = {\lambda_3 + \lambda_4\rho^2 \over 6}$, $b_{1133} = {\lambda_{S1} \over 6}$, $b_{2233} = {\lambda_{S2} \over 6}$,
$b_{1233} = - {|\lambda_{S12}|\rho \over 12}$, $b_{1123} = b_{1223} = 0$.  We have
$$f(h_1, h_2, s) \equiv \B x^4  = \sum_{i, j, k, l = 1}^3 b_{ijkl}x_ix_jx_kx_l.$$
Then $\B \in S_{4, 3}$ is a sparse fourth order three-dimensional symmetric tensor.  Forty eight of the eighty one entries of $\A$ are zero, or equivalently to say, eight of the fifteen independent entries of $\B$ are zero.  In \cite{Ka16}, a sufficient condition is presented.

As such a sparse fourth order three-dimensional symmetric tensor is special, its copositivity conditions are simpler than the conditions of Theorem \ref{t3.1}.

In particular, in (\ref{e6.15}), the powers of $x_3 = s$ only appear as $s^2$ and $s^4$.   We have
\begin{equation} \label{e6.16}
f(h_1, h_2, s) = \lambda_4s^4 + \alpha(x_1, x_2)s^2 + \beta(x_1, x_2),
\end{equation}
where
$$\alpha(x_1, x_2)= \lambda_{S1}x_1^2 - |\lambda_{S12}|\rho x_1x_2 + \lambda_{S2}x_2^2,$$
$$\beta(x_1, x_2) = \lambda_1x_1^4 + (\lambda_3 + \lambda_4\rho^2)x_1^2x_2^2 + \lambda_2x_2^4.$$

\begin{theorem} \label{t6.1}
Let $f$ be defined as above.   Then $f(h_1, h_2, s) \ge 0$ for all $h_1 \ge 0, h_2 \ge 0, s \ge 0$ if and only if the following two conditions hold.

(1) $\lambda_S \ge 0$, $\lambda_1 \ge 0$, $\lambda_2 \ge 0$, $\lambda_3 + \lambda_4\rho^2 \ge - 2\sqrt{\lambda_1\lambda_2}$, $\lambda_{S2} \ge -2\sqrt{\lambda_S\lambda_2}$;

(2) $\alpha(1, t) \ge -2\sqrt{\lambda_S\beta(1, t)}$ for all $t \ge 0$.
\end{theorem}
{\bf Proof}  In (\ref{e6.16}), regard $f$ as a quartic polynomial of $s$, which has only the terms of $s^2$ and $s^4$.   Then $f(h_1, h_2, s) \ge 0$ for all $h_1 \ge 0, h_2 \ge 0, s \ge 0$ if and only if the following two conditions hold.

(A) $\lambda_S \ge 0$, $\beta(x_1, x_2) \ge 0$ for all $x_1 \ge 0, x_2 \ge 0$;

(B) $\alpha(x_1, x_2) \ge -2\sqrt{\lambda_S\beta(x_1, x_2)}$ for all $x_1 \ge 0, x_2 \ge 0$.

We see that $\beta(x_1, x_2) \ge 0$ if and only if $\lambda_1 \ge 0$, $\lambda_2 \ge 0$ and $\lambda_3 + \lambda_4\rho^2 \ge - 2\sqrt{\lambda_1\lambda_2}$.

Discuss condition (B) in two cases.

(B1) $x_1 = 0$.   Then (B) is equivalent to $\lambda_{S2} \ge -2\sqrt{\lambda_S\lambda_2}$ in this case.

(B2) $x_1 > 0$.   Let $t = {x_2 \over x_1}$.   Then (B) is equivalent to $\alpha(1, t) \ge -2\sqrt{\lambda_S\beta(1, t)}$ for all $t \ge 0$ in this case.

Hence, conditions (A) and (B1) are equivalent to condition (1); condition (B2) is equivalent to condition (2).
\qed

If $x_1 = 0$, then (B) is equivalent to $\lambda_{S1} \ge -2\sqrt{\lambda_S\lambda_1}$ in this case.   Now this inequality is implicitly contained in condition (2).   We may add this condition to (1).   The theorem is still true.   We will do this in the statement of Theorem \ref{t5.2}.

Condition (1) of Theorem \ref{t6.1} is explicitly given.   Thus, we only need to analyze condition (2) of Theorem \ref{t6.1} further.


Now we are ready to analyze condition (2). For convenience of notation, let $g(t) \equiv 4\lambda_S\beta(1, t) - [\alpha(1, t)]^2 =b_0 t^4+b_1 t^3+b_2 t^2+b_3 t+b_4$.

(a)  If $\alpha(1,t)\geq 0$ for all $t\geq 0$, that is, $\lambda_{S1}, \lambda_{S2}\geq 0$ and $|\lambda_{S12}|\rho\leq 2\sqrt{\lambda_{S1}\lambda_{S2}}$.
Then condition (2) holds.

(b)  If $\alpha(1,t)< 0$ for all $t\geq 0$, that is, $\lambda_{S1} \le 0$ and $\lambda_{S2}\leq 0$, then
condition (2) holds if and only if the coefficients of $g$ satisfy Theorem \ref{le:24}.

(c)  Assume that $\alpha(1,t)$ is indefinite for all $t\geq 0$.
That is, there are $t_1,t_2\geq 0$ such that $\alpha(1,t_1)>0$ and $\alpha(1,t_2)<0$.
For such a case, there exist three subcases.

(i) $\lambda_{S2}=0$,  $b_0=4\lambda_S \lambda_1\geq 0$
and  $b_1=0$.  Note that this subcase we must have $|\lambda_{S12}|\rho \not = 0$.   Otherwise, we must have case (a) or (b).    Thus, we always have $\rho > 0$ in this subcase. Furthermore, $\alpha(1,t)\leq 0$ and $\rho > 0$ implies that
\[t\geq \frac{\lambda_{S1}}{|\lambda_{S12}|\rho}.\]
Together with $t\geq 0$, we need $g(t) \ge 0$ for all $t$ satisfying
\[t\geq \bar\lambda:=\max\left\{\frac{\lambda_{S1}}{|\lambda_{S12}|\rho}, 0\right\}.\]
Let $u = t - \bar \lambda$.   Then $g_1(u) = g(u - \bar \lambda)$ is a quartic polynomial of $u$. Then
condition (2) holds in this subcase if and only if the coefficients of $g_1$ satisfy Theorem \ref{le:24}.


(ii) $\lambda_{S2}>0$ and $\Delta=(\lambda_{S12}\rho)^2-4\lambda_{S1}\lambda_{S2}>0$. We need $g(t) \ge 0$ for all $t$ satisfying
\[\max\left\{\frac{|\lambda_{S12}|\rho -\sqrt{(\lambda_{S12}\rho )^2-4\lambda_{S1}\lambda_{S2}}}{2\lambda_{S2}},0\right\}:=\bar t
\leq t\leq \tilde{t}:=\frac{|\lambda_{S12}|\rho +\sqrt{(\lambda_{S12}\rho )^2-4\lambda_{S1}\lambda_{S2}}}{2\lambda_{S2}}.\]
Let $u=\frac{1}{t-\bar t}$. Then $\bar t \le t \le \tilde t$ is equivalent to $u\in [\frac{1}{\tilde{t}-\bar t},+\infty)$.   Let $g(t)=g(\frac{1}{u}+\bar t)=\frac{g_2(u)}{u^4}$.  Then $g_2(u)$ is a quartic polynomial of $u$ and the condition that $g(t) \ge 0$ for all $t \ge 0$ is equivalent to that $g_2(u)\geq 0$ for all $u\geq \frac{1}{\tilde{t}-\bar t}$.   Let $g_3(u):=g_2(u-\frac{1}{\tilde{t}-\bar t})$, which is also a quartic polynomial of $u$. Then
 condition (2) holds in this subcase if and only if the coefficients of $g_3$ satisfy Theorem \ref{le:24}.

(iii) $\lambda_{S2}<0$ and $\Delta=(\lambda_{S12}\rho)^2-4\lambda_{S1}\lambda_{S2}>0$. Let
\[\bar t_1:=\frac{|\lambda_{S12}|\rho -\sqrt{(\lambda_{S12}\rho )^2-4\lambda_{S1}\lambda_{S2}}}{2\lambda_{S2}},\quad \bar t_2:=\frac{|\lambda_{S12}|\rho +\sqrt{(\lambda_{S12}\rho )^2-4\lambda_{S1}\lambda_{S2}}}{2\lambda_{S2}}.\]

If $\bar t_1\geq 0$, then we need $g(t)\geq 0$ for all $0\leq t\leq \bar t_1$ and $t\geq \bar t_2$. For the case that $0\leq t\leq \bar t_1$, by a transformation similar to the transformation in (ii), we have
a quartic polynomial $g_4(u)$ such that condition (2) holds in this subcase if and only if the coefficients of $g_4$  satisfy Theorem \ref{le:24}. For the case that $t\geq \bar t_2$, let $g_5(u)=g(u-\bar t_2)$.  Then condition (2) holds in this subcase if and only if the coefficients of $g_5$ satisfy Theorem \ref{le:24}.

If $\bar t_1<0$, then we need $g(t)\geq 0$ for all $t\geq \bar t_2$. For such a case, let $g_6(u)=g(u-\bar t_2)$. Then $g_6(u)$ is a quartic polynomial of $u$ and
condition (2) holds in this subcase if and only if the coefficients of $g_6$ satisfy Theorem \ref{le:24}.

Thus, all the conditions of Theorem \ref{t6.1} can be analytically expressed.   We summarize the above discussion to the following theorem.

\begin{theorem} \label{t5.2}
Let $f$ be defined as above. Let $g(t) \equiv 4\lambda_S\beta(1, t) - [\alpha(1, t)]^2 =b_0 t^4+b_1 t^3+b_2 t^2+b_3 t+b_4$.  Then $f(h_1, h_2, s) \ge 0$ for all $h_1 \ge 0, h_2 \ge 0, s \ge 0$ if and only if the following two conditions hold.

(1) $\lambda_S \ge 0$, $\lambda_1 \ge 0$, $\lambda_2 \ge 0$, $\lambda_3 + \lambda_4\rho^2 \ge - 2\sqrt{\lambda_1\lambda_2}$, $\lambda_{S1} \ge -2\sqrt{\lambda_S\lambda_1}$, $\lambda_{S2} \ge -2\sqrt{\lambda_S\lambda_2}$;

(2) Either

(a) $\lambda_{S1}, \lambda_{S2}\geq 0$ and $|\lambda_{S12}|\rho\leq 2\sqrt{\lambda_{S1}\lambda_{S2}}$; or

(b) $\lambda_{S1} \le 0$, $\lambda_{S2}\leq 0$, and the coefficients of $g$ satisfy Theorem \ref{le:24}; or

(c) $\lambda_{S2}=0$,  $b_0=4\lambda_S \lambda_1\geq 0$, $b_1=0$, and the coefficients of $g_1$ satisfy Theorem \ref{le:24}, where $g_1(u) = g(u - \bar \lambda)$,
\[\bar\lambda:=\max\left\{\frac{\lambda_{S1}}{|\lambda_{S12}|\rho}, 0\right\};\]
or

(d) $\lambda_{S2}>0$, $\Delta=(\lambda_{S12}\rho)^2-4\lambda_{S1}\lambda_{S2}>0$, and the coefficients of $g_3$ satisfy Theorem \ref{le:24}, where $g_3(u):=g_2(u-\frac{1}{\tilde{t}-\bar t})$, $g_2(u) = u^4g(\frac{1}{u}+\bar t)$,
\[\bar t = \max\left\{\frac{|\lambda_{S12}|\rho -\sqrt{(\lambda_{S12}\rho )^2-4\lambda_{S1}\lambda_{S2}}}{2\lambda_{S2}},0\right\}\]
\[\tilde{t}=\frac{|\lambda_{S12}|\rho +\sqrt{(\lambda_{S12}\rho )^2-4\lambda_{S1}\lambda_{S2}}}{2\lambda_{S2}};\]
or

(e) $\lambda_{S2}<0$, $\Delta=(\lambda_{S12}\rho)^2-4\lambda_{S1}\lambda_{S2}>0$, $\bar t_1\geq 0$, and the coefficients of $g_4$ and $g_5$ satisfy Theorem \ref{le:24}, where $g_4(u):=g_7(u-\frac{1}{\bar t_2-\bar t_1})$, $g_7(u) = u^4g(\frac{1}{u}+\bar t_1)$, $g_5(u)=g(u-\bar t_2)$,
\[\bar t_1:=\frac{|\lambda_{S12}|\rho -\sqrt{(\lambda_{S12}\rho )^2-4\lambda_{S1}\lambda_{S2}}}{2\lambda_{S2}},\quad \bar t_2:=\frac{|\lambda_{S12}|\rho +\sqrt{(\lambda_{S12}\rho )^2-4\lambda_{S1}\lambda_{S2}}}{2\lambda_{S2}};\]
or

(f) $\lambda_{S2}<0$, $\Delta=(\lambda_{S12}\rho)^2-4\lambda_{S1}\lambda_{S2}>0$, $\bar t_1 < 0$, the coefficients of $g_6$ satisfy Theorem \ref{le:24}, where $g_6(u)=g(u-\bar t_2)$, $\bar t_1$ and $\bar t_2$ are defined as above.
\end{theorem}

\bigskip


\end{document}